\documentclass[prl,twocolumn,showpacs,preprintnumbers,superscriptaddress]{revtex4}

\usepackage{amsmath}
\usepackage{amssymb}
\usepackage{graphicx}

\newcommand{\sgn}{\mathop{\rm sign}\nolimits}
\newcommand{\diag}{\mathop{\rm diag}\nolimits}
\newcommand{\ev}{\mathop{\rm ev}\nolimits}
\newcommand{\spann}{\mathop{\rm span}\nolimits}
\newcommand{\Tr}{\mathop{\rm Tr}\nolimits}
\newcommand{\aCom}[2]{{\{ #1, #2 \}}}
\newcommand{\Com}[2]{{[ #1, #2 ]}}
\newcommand{\dg}{\dagger}

\newcommand{\rv}{{\mathrm v}}

\newcommand{\hti}{{\tilde h}}
\newcommand{\fti}{{\tilde f}}

\newcommand{\ga}{{\alpha}}
\newcommand{\go}{{\omega}}
\newcommand{\gx}{{\chi}}

\newcommand{\gd}{{\delta}}

\newcommand{\gk}{{\kappa}}
\newcommand{\gt}{{\tau}}

\newcommand{\gre}{{\cal E}}

\newcommand{\vG}{{\check G}}

\newcommand{\vgt}{{\check \tau}}
\newcommand{\vOne}{{\check 1}}

\newcommand{\vcA}{{\check{\cal A}}}
\newcommand{\vcB}{{\check{\cal B}}}

\newcommand{\cT}{{T_e}}

\newcommand{\Vb}{{\bar V}}
\newcommand{\dV}{{\Delta V}}

\newcommand{\matr}[1]{\bigl (
                        \begin{smallmatrix}
                        #1
                        \end{smallmatrix}
                      \bigr )}

\begin{document}


\title{Elementary events of electron transfer
in a voltage-driven quantum point contact}

\author{Mihajlo Vanevi\' c}%
\affiliation{Departement f\" ur Physik und Astronomie,
Klingelbergstrasse 82, 4056 Basel, Switzerland}

\author{Yuli V. Nazarov}
\affiliation{Kavli Institute of Nanoscience, Delft University of
Technology, 2628 CJ Delft, The Netherlands}

\author{Wolfgang Belzig}
\affiliation{Fachbereich Physik, Universit\" at Konstanz, D-78457
Konstanz, Germany}

\date{\today}

\begin{abstract}
  We find that the statistics of electron transfer in a coherent
  quantum point contact driven by an arbitrary time-dependent
  voltage is composed of elementary events of two kinds:
  unidirectional one-electron transfers determining the average
  current and bidirectional two-electron processes contributing to
  the noise only. This result pertains at vanishing temperature
  while the extended Keldysh-Green's function formalism in use
  also enables the systematic calculation of the higher-order
  current correlators at finite temperatures.
\end{abstract}




\maketitle

The most detailed description of the charge transfer in coherent
conductors is a statistical one.  At constant bias, the full
counting statistics (FCS) of electron transfer
\cite{art:LevitovLesovikJETP93} can be directly interpreted in
terms of elementary events independent at different energies. The
FCS approach is readily generalized to the case of a
time-dependent voltage bias
\cite{art:IvanovLevitovJETP93,art:LevitovJMath}. The current
fluctuations in coherent systems driven by a periodic voltage
strongly depend on the shape of the driving
\cite{art:LevitovNonstatAB}, which frequently is not apparent in
the average current \cite{art:Pedersen}. The noise power, for
instance, exhibits at low temperatures a piecewise linear
dependence on the dc voltage with kinks corresponding to integer
multiples of the ac drive frequency and slopes which depend on the
shape and the amplitude of the ac component. This dependence has
been observed experimentally in normal coherent conductors
\cite{art:SchoelkopfPRL98} and diffusive normal
metal--superconductor junctions \cite{art:KozhevnikovPRL00}.

The elementary events of charge transfer driven by a general
time-dependent voltage have not been identified so far. The time
dependence mixes the electron states at different energies
\cite{art:Pedersen} which makes this question both interesting and
non-trivial. The first step in this research has been made in
\cite{art:LeeLevitovCMat95} for a special choice of the
time-dependent voltage. The authors have considered a
superposition of overlapping Lorentzian pulses of the same sign
("solitons"), with each pulse carrying a single charge quantum.
The resulting charge transfer is unidirectional with a binomial
distribution of transmitted charges. The number of attempts per
unit time for quasiparticles to traverse the junction is given
by the dc component of the voltage, independent of the overlap
between the pulses and their duration
\cite{art:IvanovLeelevitovPRB97}. It has been shown that such
superposition minimizes the noise reducing it to that of a
corresponding dc bias.
A microscopic picture behind the soliton pulses has been revealed
only recently \cite{art:KlichLevitov06}. In contrast to a general
voltage pulse which can in principle create a random number of
electron-hole pairs with random directions, a soliton pulse at
zero temperature always creates a single electron-hole pair with
quasiparticles moving in opposite directions. One of the
quasiparticles (say, electron) comes to the contact and takes part
in the transport while the hole goes away. Therefore, soliton
pulses can be used to create minimal excitation states with "pure"
electrons excited from the filled Fermi sea and no holes left
below. The existence of such states can be probed by noise
measurements
\cite{art:KlichLevitov06,art:ButtikerPRB05,art:MuzykantskiiPRB05}.
%

\begin{figure}[b]
\includegraphics[width=8.5cm, height=2.62cm]{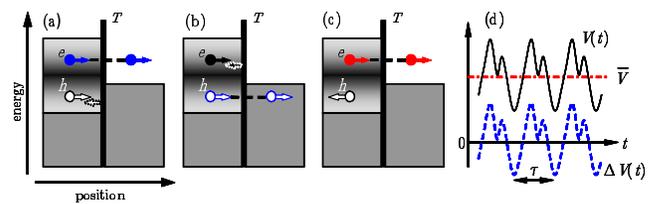}
\caption{\label{fig:Scheme} Schematic representation of elementary
    events: bidirectional (a, b) and unidirectional (c).
    Shifts of the effective chemical potential
    in the left lead due to time-dependent voltage drive
    are indicated by shading.
    For periodic drive, the dc voltage component [panel (d),
    dash-dotted line] describes unidirectional charge transfer,
    while the ac component (dashed curve) describes bidirectional
    events affecting the noise and higher-order even cumulants. }
\end{figure}
%
In this Letter, we do identify the independent elementary events
for an arbitrary time-dependent driving applied to a generic
conductor. Since generic conductor at low energies can be
represented as a collection of independent transport channels, it
is enough to specify elementary events for a single channel of
transmission $T$. The answer is surprisingly simple. There are two
kinds of such events: We call them {\it bidirectional} and {\it
unidirectional}. In the course of a {\it bidirectional} event $k$
an electron-hole pair is created with probability
$\sin^2(\ga_k/2)$, with $\alpha_k$ being determined by the details
of the time-dependent voltage. The electron and hole move in the
{\it same} direction reaching the scatterer. The charge transfer
occurs if the electron is transmitted and the hole is reflected,
or vice versa [Fig. \ref{fig:Scheme}(a,b)]. The probabilities of
both outcomes, $T R$ ($R$ being reflection coefficient), are the
same. Therefore, the bidirectional events do not contribute to the
average current and odd cumulants of the charge transferred
although they do contribute to the noise and higher-order even
cumulants. A specific example of a bidirectional event for a
soliton-antisoliton pulse was given in
\cite{art:IvanovLeelevitovPRB97}.

The {\it unidirectional} events are the same as for a constant
bias or a soliton pulse. They are characterized by chirality
$\gk_l=\pm 1$ which gives the direction of the charge transfer. An
electron-hole pair is always created in the course of the event,
with electron and hole moving in opposite directions
[Fig.~\ref{fig:Scheme}(c)]. Either electron ($\gk_l=1$) or hole
($\gk_l=-1$) passes the contact with probability $T$, thus
contributing to the current.

Mathematically, the above description corresponds to the cumulant
generating function $S(\gx)=S_1(\gx)+S_2(\gx)$, where
\begin{equation}\label{eq:Stypical}
S_1
=
2\sum_k \ln\left[1+T R \sin^2\left(\frac{\ga_k}{2}\right)
(e^{i\gx}+e^{-i\gx}-2)\right]
\end{equation}
and
\begin{equation}\label{eq:Sspec}
S_2 = 2 \sum_l
\ln[ 1 + T (e^{-i \gk_l \gx}-1)]
\end{equation}
are the contributions of the bidirectional and unidirectional
events, respectively. Here $\gx$ is the counting field, and
$\ga_k$ and $\gk_l$ are the parameters of the driving to be
specified later. The sum in both formulas is over the set of
corresponding events \cite{countability}. The elementary events
have been inferred from the form of the cumulant-generating
function, as it has been done in~\cite{art:events,art:events1}.

The cumulant-generating function
given by Eqs.~\eqref{eq:Stypical} and \eqref{eq:Sspec}, together
with the interpretation, is the main result of this Letter. It
holds at zero temperature only: since the elementary events are
the electron-hole pairs created by the applied voltage, the
presence of thermally excited pairs will smear the picture.
Equations ~\eqref{eq:Stypical} and \eqref{eq:Sspec} contain the
complete $\gx$-field dependence in explicit form which allows for
the calculation of higher-order cumulants and charge transfer
statistics for arbitrary time-dependent voltage. The probability
that $N$ charges are transmitted within the time of measurement is
given by $P(N)=(2\pi)^{-1}\int_{-\pi}^\pi d\gx \exp[S(\gx) -
iN\gx]$. Higher-order derivatives of $S$ with respect to $\gx$ are
proportional to the cumulants of transmitted charge, or
equivalently, to higher-order current correlators at zero
frequency. The details of the driving are contained in the set of
parameters $\{\ga_k\}$ and separated from the $\gx$-field
dependence. This opens an interesting possibility to excite the
specific elementary processes and design the charge transfer
statistics by appropriate time dependence of the applied voltage,
with possible applications in production and detection of the
many-body entangled states
\cite{art:events,art:BlatterPRB05,art:Beenakker}.

Below we present the microscopic derivation of
Eqs.~\eqref{eq:Stypical} and \eqref{eq:Sspec}. We neglect charging
effects and assume instantaneous scattering at the contact with
quasiparticle dwell times much smaller than the characteristic
time scale of the voltage variations.
The approach we use is the nonequilibrium Keldysh-Green's function
technique, extended to access the full counting statistics
\cite{art:NazarovAnnPhys,art:BelzigPRL01,art:BelzigInBook,%
art:NazarovEPJB03}.
The Green's functions of the left ($1$) and right ($2$)
leads are given by \cite{art:BelzigPRL01,art:BelzigInBook}
\begin{equation}
\vG_1 = e^{-i\gx\vgt_{1}/2}
    \begin{pmatrix}
    1 & 2\hti \\
    0  & -1
    \end{pmatrix}
e^{i\gx\vgt_{1}/2},
\quad
\vG_2 =
    \begin{pmatrix}
    1 & 2h \\
    0 & -1
    \end{pmatrix},
\end{equation}
where $\vgt_1=\matr{ 0 & 1\\ 1 & 0}$ is a matrix in
Keldysh($\,\check{}\,$) space.
Hereafter we use a compact operator notation in which
the time (or energy) indices are suppressed and the products are
interpreted in terms of convolution over internal indices,
e.g.,
$(\vG_1\vG_2)(t',t'')=\int dt_1\vG_1(t',t_1)\vG_2(t_1,t'')$
(and similar in the energy representation).
The equilibrium Green's function $\vG_2(t'-t'')$ depends only
on time difference. In the energy representation
$\vG_2(\gre',\gre'')$ is diagonal in energy indices
with $h(\gre',\gre'')=\tanh(\gre'/2\cT)\; 2\pi\gd(\gre'-\gre'')$.
Here the quasiparticle energy $\gre$ is measured with respect to the
chemical potential in the absence of the bias and $\cT$ is the
temperature. The Green's function $\vG_1(t',t'')$ depends on two
time (or energy) arguments. It takes into account the effect of
applied voltage $V(t)$ across the junction through the gauge
transformation $\hti = UhU^\dg$ which makes $\vG_1$ nondiagonal
in energy representation. The unitary operator $U$ is given by
$ U(t',t'') = f(t')\gd(t'-t'')$ in the time representation,
where $ f(t') = \exp[-i \int_0^{t'} eV(t) dt]$.
The cumulant generating function $S(\gx)$ of the charge transfer
through the junction is given
by \cite{art:BelzigInBook,art:NazarovSuperlatt99}
\begin{equation}\label{eq:S1}
S(\gx) =
\Tr \ln
    \left [
    \vOne + \frac{T}{2}
        \left(
        \frac{\aCom{\vG_1}{\vG_2}}{2} - \vOne
        \right)
    \right ].
\end{equation}
Here the trace and products of Green's functions include both
summation in Keldysh indices and integration over time (energy).
For a dc voltage bias, $\vG_1$ and $\vG_2$ are diagonal
in energy indices and $S(\gx)$ is readily interpreted in terms of
elementary events independent at different
energies~\cite{art:BelzigInBook}.
To deduce the elementary events in the presence of time
dependent voltage drive it is necessary to diagonalize
$\aCom{\vG_1}{\vG_2}_{\gre'\gre''}$. The diagonalization procedure
is described in the following.

For the anticommutator of the Green's functions we find
$\aCom{\vG_1}{\vG_2}/2-\vOne = -2 \sin(\gx/2)(\vcA+\vcB)$,
with
$
\vcA =
    \matr{
    1 & b \\
    0 & 0}
    \otimes A
$,
and
$
\vcB =
    \matr{
    0 & -b \\
    0 & 1}
    \otimes B
$.
Here
$A = (1-h\hti)\sin(\gx/2) + i (h-\hti)\cos(\gx/2)$,
$B = (1-\hti h)\sin(\gx/2) + i (h-\hti)\cos(\gx/2)$,
$b=-i \cot(\gx/2)$, and $\otimes$ is the tensor product.
Since $\vcA\vcB = \vcB\vcA = 0$, the operators $\vcA$ and $\vcB$
commute and satisfy for integer $n$:
$
(\vcA + \vcB)^n
=
    \matr{
    1 & b \\
    0 & 0}
    \otimes A^n
+
    \matr{
    0 & -b \\
    0 & 1}
    \otimes B^n
$.
Therefore, $S(\gx)$ given by
Eq.~\eqref{eq:S1} reduces to
\begin{equation}\label{eq:S2}
S =
\Tr\ln\left[1-T \sin\left(\frac{\gx}{2}\right)A\right]
+
\Tr\ln\left[1-T \sin\left(\frac{\gx}{2}\right)B\right].
\end{equation}
%
A further simplification of $S(\gx)$ is possible in the zero
temperature limit, in which the hermitian $h$-operators are
involutive: $h^2 = \hti^2 = 1$. The operators $h\hti$ and $\hti h$
are mutually inverse and commute with each other. Because $h\hti$
is unitary, it has the eigenvalues of the form $e^{i\ga_k}$ with
real $\ga_k$, and possesses an orthonormal eigenbasis
$\{\rv_{\ga_k}\}$. The {\it typical} eigenvalues of $h\hti$ (or
$\hti h$) appear in pairs $e^{\pm i\ga}$ with the corresponding
eigenvectors $\rv_\ga$ and $\rv_{-\ga}=h\rv_\ga$. In the basis
$(\rv_\ga,\rv_{-\ga})$ operators $h\hti$ and $\hti h$ are
diagonal and given by $h\hti = \diag(e^{i\ga},e^{-i\ga})$ and
$\hti h = \diag(e^{-i\ga},e^{i\ga})$.
The eigensubspaces $\spann(\rv_\ga,\rv_{-\ga})$ of the
anticommutator $\aCom{h}{\hti}$ are invariant with respect to
$h$, $\hti$, and $A$ because of
$\Com{h}{\aCom{h}{\hti}}=\Com{\hti}{\aCom{h}{\hti}}=0$.
The operators $h$ and $\hti$ are anti-diagonal in the basis
$(\rv_\ga,\rv_{-\ga})$, with matrix components
$h_{12}=1$, $h_{21}=1$, $\hti_{12}=e^{-i \ga}$, and
$\hti_{21}=e^{i \ga}$. The operator $A$ can be
diagonalized in invariant subspaces, with typical eigenvalues
given by
\begin{multline}\label{eq:evA}
\ev A = 2\sin(\ga/2)
\bigg(
\sin(\ga/2)\sin(\gx/2)
\\
\pm i \sqrt{1-\sin^2(\ga/2)\sin^2(\gx/2)}
\bigg).
\end{multline}
Similarly, we obtain $\ev B = \ev A$.
From Eqs.~\eqref{eq:S2} and \eqref{eq:evA} we recover the
generating function $S_1$ given by Eq.~\eqref{eq:Stypical},
which is associated with the paired eigenvalues $e^{\pm i\ga_k}$.

There are, however, some {\it special} eigenvectors of $h\hti$
which do not appear in pairs. The pair property discussed
above was based on the assumption that $\rv_\ga$ and $h \rv_\ga=
\rv_{-\ga}$ are linearly independent vectors. In the special case,
these vectors are the same apart from a coefficient. Therefore,
the special eigenvectors of $h\hti$ are the eigenvectors of both $h$
and $\hti$ with eigenvalues ${\pm 1}$. This means that the special
eigenvectors posses {\it chirality}, with positive (negative)
chirality defined by $h\rv=\rv$ and $\hti \rv = -\rv$ ($h\rv=-\rv$
and $\hti \rv = \rv$).
From Eq.~\eqref{eq:S2} we obtain the generating function
$S_2(\gx)$ given by Eq.~\eqref{eq:Sspec}, where $l$ labels the
special eigenvectors and $\gk_l$ is the chirality.

In the following we focus on a periodic driving $V(t+\gt)=V(t)$
with the period $\gt=2\pi/\go$, for which the eigenvalues of
$h\hti$ can be easily obtained by matrix diagonalization. The
operator $\hti$ couples only energies which differ by an integer
multiple of $\go$, which allows to map the problem into the energy
interval $0<\gre<\go$ while retaining the discrete matrix
structure in steps of $\go$. Therefore, the trace operation in Eq.
\eqref{eq:S1} becomes an integral over $\gre$ and the trace in
discrete matrix indices. The operator $h\hti$ in the energy
representation is given by
$(h\hti)_{nm}(\gre)
=
\sgn(\gre+n\go)
\sum_k \fti_{n+k}\fti_{m+k}^*
\sgn(\gre-k\go-e\Vb)$,
with
$\fti_n
=
(1/\gt) \int_{-\gt/2}^{\gt/2}dt\;
e^{-i\int_0^t dt'\; e\dV(t')}  e^{in\go t}$.
Here $\Vb = (1/\gt)\int V(t)dt$ is the dc voltage offset and
$\dV(t) = V(t) - \Vb$ is the ac voltage component. The
coefficients $\fti_n$ satisfy $\sum_k \fti_{n+k}\fti_{m+k}^* =
\gd_{nm}$ and $\sum_n n |\fti_n|^2 = 0$.

To evaluate $S(\gx)$ for a given periodic voltage drive $V(t)$ it
is necessary to diagonalize $(h\hti)_{nm}(\gre)$. First we analyze
the contribution of typical eigenvalues $e^{\pm i\ga}$.
The matrix $(h\hti)_{nm}(\gre)$ is piecewise constant for
$\gre\in(0,\go_1)$ and $\gre\in(\go_1,\go)$, where $\go_1 = e\Vb -
N \go$ and $N=\lfloor e\Vb/\go \rfloor$ is the largest integer
less than or equal $e\Vb/\go$. The eigenvalues $e^{\pm
i\ga_{kL(R)}}$ of $(h\hti)_{nm}$ are calculated
for $\gre\in(0,\go_1)$ [$\gre\in(\go_1,\go)$] using
finite-dimensional matrices, with the cutoff in
indices $n$ and $m$ being much larger than the characteristic
scale on which $|\fti_n|$ vanish. Further increase of the size of
matrix just brings more eigenvalues with $\ga_k=0$ which do not
contribute to $S(\gx)$, and does not change the rest
with $\ga_k\ne 0$. This is a
signature that all important Fourier components of the drive are
taken into account. The eigenvalues $e^{\pm i\ga_{kL(R)}}$ give
rise to two terms, $S_1=S_{1L}+S_{1R}$, with
\begin{multline}\label{eq:S1LRPer}
S_{1L,R}(\gx)
=
M_{L,R}
\sum_{k}
\ln[
1+TR \sin^2(\ga_{kL,R}/2)
\\
\times
(e^{i\gx}+e^{-i\gx}-2)
].
\end{multline}
Here $M_L = t_0\go_1/\pi$, $M_R=t_0(\go-\go_1)/\pi$, and $t_0$ is
the total measurement time which is much larger than $\gt$ and the
characteristic time scale on which the current fluctuations are
correlated.

The special eigenvectors all have the same chirality which is
given by the sign of the dc offset $\Vb$. For $e\Vb>0$,
there are $N_1 = N+1$ special eigenvectors for $\gre\in(0,\go_1)$
and $N_2=N$ for $\gre\in(\go_1,\go)$. Because $e\Vb = N_1\go_1 +
N_2(\go-\go_1)$, the effect of the special eigenvectors is the
same as of the dc bias
\begin{equation}\label{eq:S2Per}
S_2(\gx) = \frac{t_0e\Vb}{\pi} \ln[1+T(e^{-i\gx}-1)].
\end{equation}
Comparing Eqs.~\eqref{eq:Sspec} and \eqref{eq:S2Per} we see that
unidirectional events for periodic drive are uncountable. The
summation in Eq.~\eqref{eq:Sspec} stands both for the energy
integration in the interval $\go$ and the trace in the discrete
matrix indices. In the limit of a single pulse $\go\to 0$
unidirectional events remain uncountable for a generic voltage,
while being countable, e.g., for soliton pulses carrying
integer number of charge
quanta~\cite{art:IvanovLeelevitovPRB97}.

Equations~\eqref{eq:S1LRPer} and \eqref{eq:S2Per} determine the
charge transfer statistics at zero temperature for an arbitrary
periodic voltage applied. The generating function consists of a
binomial part ($S_2$) which depends on the dc offset $\Vb$
only, and a contribution of the ac voltage component ($S_1$)
[Fig.~\ref{fig:Scheme}(d)]. The latter is the sum of two terms
which depend on the number of unidirectional attempts per period
$e\Vb/\go$. The simplest statistics is obtained for an integer
number of attempts for which $S_{1L}$ vanishes
\cite{art:IvanovLevitovJETP93}. The Fourier components of
the optimal Lorentzian pulses
$ V_L(t) = (2\gt_L/e) \sum_k
[(t-k\gt)^2+\gt_L^2]^{-1}$
of width $\gt_L>0$ are given by $\fti_{-1}=-e^{-2\pi\gt_L/\gt}$,
$\fti_n = e^{-2\pi n\gt_L/\gt}-e^{-2\pi (n+2)\gt_L/\gt}$
for $n\ge 0$, and $\fti_n = 0$ otherwise.
In this case $S_{1L}=S_{1R}=0$ and the
statistics is {\it exactly} binomial with one electron-hole
excitation per period, in agreement with Refs.
\cite{art:IvanovLeelevitovPRB97,art:KlichLevitov06}.

\begin{figure}[tb]
\includegraphics[width=7.3cm, height=4.8cm]{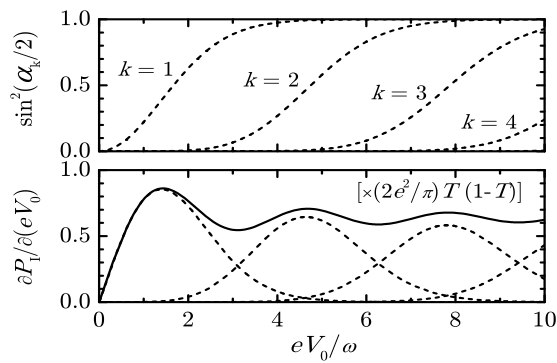}
\caption{\label{fig:Noise} The probability of elementary events
   for harmonic drive with amplitude $V_0$ (upper panel).
   With increasing amplitude more and more eigenvalues $\ga_k$ come
   into play and contribute to transport. The derivative of the
   noise power with respect to $V_0$ (---) decomposed into
   contributions (-- --) of elementary events (lower panel).}
\end{figure}
%
The elementary events at zero temperature can be probed by noise
measurements. For example, in the case of an ac drive with
$\Vb=0$, only bidirectional events of $R$-type remain
[$S(\gx)=S_{1R}(\gx)$]. Both the number of events and their
probabilities increase with increasing the driving amplitude
$V_0$, which results in the characteristic oscillatory change of
the slope of the current noise power $P_I = (4e^2\go/\pi)T(1-T)
\sum_k \sin^2(\ga_{k}/2)$. The decomposition of $\partial
P_I/\partial V_0$ into contributions of elementary events for
harmonic drive is shown in Fig.~\ref{fig:Noise}.

Our method also enables the efficient and systematic analytic
calculation of the higher-order cumulants at finite temperatures.
They can be obtained directly from Eq.~\eqref{eq:S2} by
expansion in the counting field to the certain order
before taking the trace. The trace of a finite number of terms can
be taken in the original basis in which $\vG_1$ and $\vG_2$ are
defined. The details of this approach will be given
elsewhere. However, the formulas obtained (as a function of
$\{\fti_n\}$) can not be interpreted as elementary events term by
term. To identify the elementary events it is necessary to find
$S(\gx)$ which requires full expansion or diagonalization, as
presented above.

In conclusion, we have studied the statistics of the charge
transfer in a quantum point contact driven by time-dependent
voltage. We have deduced the elementary transport processes at zero
temperature from an analytical result for the cumulant
generating function. The transport consists of unidirectional
and bidirectional charge transfer events which can be interpreted in
terms of electrons and holes which move in opposite and the same
directions, respectively. Unidirectional events account for the net
charge transfer and are described by binomial cumulant generating
function which depends on the dc voltage offset. Bidirectional
events contribute only to even cumulants of charge transfer at
zero temperature. They are created with probability which depends on
the shape of the ac voltage component.
The statistics of charge transfer is the simplest for an integer
number of attempts for quasiparticles to traverse the junction.
This results in photon-assisted effects in even-order cumulants as
a function of a dc offset. The approach we have used also allows
for the systematic calculation of higher-order cumulants at finite
temperatures.

We acknowledge valuable discussions with L. S. Levitov and C.
Bruder. This work has been supported by the Swiss NSF and NCCR
"Nanoscience" (MV), the DFG through SFB 513 and the Landesstiftung
Baden-W\"urttemberg (WB).


\begin{thebibliography}{}
\bibitem{countability}
The set of the unidirectional events is not necessary countable,
see below.
\end{thebibliography}
\end{document}